\documentclass[12pt]{article}

\usepackage{jcappub1}
\usepackage{amsmath}
\usepackage{graphicx}
\usepackage{verbatim}
\usepackage{xfrac}

\newcommand{\be}{\begin{equation}}
\newcommand{\ee}{\end{equation}}

\newcommand{\ba}{\begin{eqnarray}}
\newcommand{\ea}{\end{eqnarray}}
\newcommand{\bs}{\begin{subequations}}
\newcommand{\es}{\end{subequations}}

\title{Constraining symmetron fields with atom interferometry}

\author[a]{Clare Burrage}
\author[a]{Andrew Kuribayashi-Coleman}
\author[a]{James Stevenson}
\author[a]{and Ben Thrussell}

\emailAdd{Clare.Burrage@nottingham.ac.uk}
\emailAdd{James.Stevenson@nottingham.ac.uk}

\affiliation[a]{School of Physics and Astronomy,\\ University of Nottingham,\\ Nottingham, NG7 2RD,\\
United Kingdom }

\abstract{We apply the new constraints from atom-interferometry searches for screening mechanisms to the symmetron model, finding that these experiments exclude a previously unexplored region of parameter space. We discuss the possibility of networks of domain walls forming in the vacuum chamber, and how this could be used to discriminate between models of screening.}

\begin{document}
\maketitle
\flushbottom

\section{Introduction} 
Theories of dark energy that introduce new, light scalar fields coupled to matter have inspired the study of screening mechanisms to explain why the associated fifth forces have not yet been detected \cite{Joyce:2014kja,Clifton:2011jh}. Screening mechanisms allow the scalar field theory to have non-trivial self-interactions, and so the properties of the scalar, and the resulting fifth force, can vary with the environment. Whilst screening mechanisms were introduced in order to explain the absence of an observation of a fifth force to date, that does not mean that such fifth forces are intrinsically unobservable. Experimental searches need only to be carefully designed to take advantage of the non-linear screening behaviour. 

Given a background field profile, the self interactions of the screened scalar field can have three possible consequences on the properties of scalar fluctuations on top of that background \cite{Joyce:2014kja}:
\begin{enumerate}
\item The mass of the fluctuations becomes dependent on the background. If the field becomes heavy in dense environments and light in diffuse ones, this can explain why the scalar force would not be detected around the macroscopic dense sources used in current fifth force experiments. This is known as the chameleon mechanism after the archetypal chameleon model \cite{Khoury:2003rn,Khoury:2003aq}. 

\item The strength of the coupling to matter becomes dependent on the background. If the field becomes weakly coupled in experimental environments it is clear that it will be harder to detect. Examples of models that employ this mechanism include the symmetron \cite{Hinterbichler:2010es,Hinterbichler:2011ca} and density dependent dilaton \cite{Damour:1994zq,Brax:2010gi}. 

\item The coefficient of the scalar kinetic term becomes dependent on the background. If the coefficient becomes large in experimental searches it becomes difficult for the scalar to propagate, and so the force is suppressed.  This effect occurs in any model which has gradient self interactions, including Galileon \cite{Nicolis:2008in} and k-essence models \cite{Babichev:2009ee,Brax:2012jr,Burrage:2014uwa}, and is called the Vainshtein mechanism \cite{Vainshtein:1972sx}.
\end{enumerate}

It has recently been demonstrated that atomic nuclei inside a high quality vacuum chamber are very sensitive probes of chameleon screening, this is because the nucleus is so small that the screening cannot work efficiently.  Forces on individual atoms can now be measured to a very high precision using atom interferometry, and as a result new constraints on chameleon models have been derived. Further improvements to these experiments are cureently underway. 

It remains to be determined whether the power of atom interferometry can be extended to constrain theories which screen through other means. Vainshtein screening will not be accessible, because the gradient self-interactions mean that the screening takes place over much longer distance scales than are achievable in a terrestrial laboratory.  In contrast, however, theories which screen by varying their coupling constant with the environment are phenomenologically similar to chameleon models, and so it is expected that atom interferometry could also provide useful constraints. 

In this work we will focus on the symmetron model, as an example of a theory which screens by varying its coupling constant.  This model is chosen because it has been shown that the model can be constructed in such a way that it is radiatively stable and quantum corrections remain under control \cite{Burrage:2016xzz}. Earlier work studied a similar model but with a different motivation~\cite{Pietroni:2005pv,Olive:2007aj}, and string-inspired models with similar phenomenology have also been proposed~\cite{Damour:1994zq,Brax:2011ja}.

In Section \ref{sec:symm} we will review the symmetron model, and how the force between two extended objects can be screened. In Section \ref{sec:atom} we apply the results of existing atom interferometry experiments to find new constraints on the symmetron model which are presented in Figure \ref{fig:constraints}.  In Section \ref{sec:domain} we discuss the possibility that domain walls could form inside the vacuum chamber, leading to the possibility that atoms could experience a symmetron force, even in the absence of a source inside the vacuum chamber. We conclude in Section \ref{sec:conclusions}.

\section{The Symmetron}
\label{sec:symm}
The simplest version of the symmetron model is as a canonical scalar field with potential 
\begin{equation}
V(\phi) = \frac{\lambda}{4} \phi^4 -\frac{\mu^2}{2}\phi^2\;,
\end{equation}
where $\lambda$ (which is dimensionless) and $\mu$ (which has mass dimensions) are the parameters of the theory which must be determined by experiment. 
The scalar field couples to matter through dimension six terms in the Lagrangian of the form
\begin{equation}
\mathcal{L} \supset \frac{\phi^2}{2 M^2}T^{\mu}_{\mu}\;,
\end{equation}
where $T_{\mu\nu}$ is the energy-momentum tensor of all of the matter fields and $M$ is an energy scale which controls the strength of the coupling to matter. 

The interactions with matter mean that in the presence of a non-relativistic, static background matter density $\rho$ the symmetron field moves in an effective potential 
\begin{equation}
V_{\rm eff}(\phi) = \frac{1}{2}\left(\frac{\rho}{M^2}-\mu^2\right)\phi^2 +\frac{\lambda}{4}\phi^4\;,
\end{equation}
from which it can be seen that when the density is sufficiently high, $\rho>M^2 \mu^2$, the effective potential has only one minimum and the field is trapped at $\phi=0$.  As the density is decreased the potential undergoes a symmetry breaking transition, and the field can roll into one of two minima with $\phi^2 = (\mu^2 -\rho/M^2)/\lambda$.

In Ref.~\cite{Hinterbichler:2011ca}, the symmetry-breaking scale is chosen close to the cosmological density today, i.e.~$\mu^2 M^2\sim H_0^2M_{\rm Pl}^2$, where $H_0$ is the present-day Hubble scale. In addition, the symmetron force in vacuum is required to have approximately gravitational strength, i.e.~$\phi/M^2 \sim 1/M_{\rm Pl}$, such that there may be observable consequences without fine-tuning of the coupling scale. However, other choices of parameters are possible.  In particular it has been shown that both E\"{o}t-Wash experiments \cite{Upadhye:2012rc}, and measurements of exo-planets \cite{Santos:2016rdg} constrain a very different region of parameter space with coupling constants $M\lesssim 10 \mbox{ TeV}$ and the mass scale $\mu$ around the electronvolt scale.  The self coupling parameter $\lambda$ is very poorly constrained currently. 

The radiatively stable model derived in \cite{Burrage:2016xzz} has a slightly more complicated potential 
\begin{equation}
V(\phi) =\left(\frac{\lambda}{16\pi}\right)^2 \phi^4 \left(\ln \frac{\phi^2}{m^2}-Y\right)\;,
\end{equation}
for constant $m$, $\lambda$ and $Y$.  However, the structure of the symmetry breaking transition, and the resulting phenomenology remains essentially the same as the original symmetron model, and so we will focus our attention on the simpler model in what follows. 

Around a static, spherically symmetric object of density $\rho_{\rm in}$ and radius $R$, embedded in a background density $\rho_{\rm out}$, the symmetron field profile is 
\begin{equation}
\phi=\phi_{\rm out} -\frac{(\phi_{\rm out}-\phi_{\rm in})R e^{m_{\rm out}(R-r)}}{r}\left(\frac{m_{\rm in}R-\tanh m_{\rm in} R}{m_{\rm in}R+Rm_{\rm out}\tanh m_{\rm in}R}\right)\;,
\end{equation}
where $m_{\rm in}$ and $m_{\rm out}$ are respectively the mass of the field inside the source object, and outside, and $\phi_{\rm in}$ and $\phi_{\rm out}$ are the values of the scalar field that minimize the effective potential inside and outside the source. 
The force on a test particle moving on top of this field profile is then given by $F= \phi\nabla \phi/M^2$.

If everywhere in the experiment the density is higher than that required for the symmetry breaking transition, $(\rho_{\rm out},\rho_{\rm in})>M^2 \mu^2$, then the field will be constrained to be $\phi=0$ everywhere and it will never be possible to see the associated fifth forces. 
If $\rho_{\rm out}< \mu^2 M^2$ then the field has a non-trivial profile inside the vacuum chamber, and the value of the field at the center of the vacuum chamber (assumed to be spherical for simplicity) depends on the relative sizes of the radius of the vacuum chamber $L$ and the Compton wavelength of the field. If $m_{\rm out} L\gtrsim1$ then 
\begin{equation}
\phi_{\rm out} =\frac{1}{\sqrt{\lambda}}\left(\mu^2 -\frac{\rho_{\rm out}}{M^2}\right)^{1/2}\;,
\end{equation}
otherwise the field does not have room to evolve away from $\phi=0$ in the walls of the chamber and we have
\begin{equation}
\phi_{\rm out}=0\;.
\end{equation}

If $\rho_{\rm in}< \mu^2 M^2$, then the source object causes only a small perturbation of the background field. In this case there is no screening, and the force has the usual Yukawa form.  On the other hand, if $\rho_{\rm in}> \mu^2 M^2$ and $m_{\rm in}R \gg 1$, then the symmetry is restored inside the source and the resulting force on a test particle is suppressed \cite{Hinterbichler:2010es}. 

We are not interested, however, in the force on an infinitesimal test particle but instead in the force between two extended objects either or both of which could be screened. Following the arguments of \cite{Hui:2009kc} if we assume a hierarchy between mass A and mass B, so that the field profile due to B can be considered a small perturbation on the field profile of A {\it at the surface of mass B}, then the force can be  found by considering the change in momentum of ball B and using the Bianchi identity. The symmetron force between two objects A and B is therefore
\begin{equation}
F_{\rm symm}=4 \pi \lambda_A\lambda_B (1+m_{\rm out}R_B)(1+m_{\rm out}r)\frac{e^{m_{\rm out}(R_A-r)}}{r^2}\;,
\end{equation}
where 
\begin{equation}
\lambda_i = \left.(\phi_{\rm out}-\phi_{\rm in})R\left(\frac{m_{\rm in}R-\tanh m_{\rm in} R}{m_{\rm in}R+Rm_{\rm out}\tanh m_{\rm in}R}\right)\right|_{i}\;,
\label{eq:lambda}
\end{equation}
where all of the quantities on the right hand side of Equation (\ref{eq:lambda}) are evaluated for the object in question.  $\lambda_i$ can therefore be considered the symmetron `charge' for the object.  The fact that we are treating the field profiles due to A and B hierarchically explains the slight asymmetry in the dependencies on $R_A$ and $R_B$.
If $\lambda_i \ll 1$ then we say that the object is screened, and the force is suppressed. 

\section{Atom Interferometry}
\label{sec:atom}
Atom interferometry has been shown to be a powerful technique for constraining chameleon models with screening \cite{Burrage:2014oza}. These experiments work by putting an atom into a superposition of states which travel on two different paths.  If the wavefunction accumulates a phase difference between the two paths this can be detected as an interference pattern when the two path are merged \cite{Storey:1994oka,feynmanhibbs}.  If the atoms experience a constant acceleration in the same direction as the separation between the paths then this results in exactly such a phase difference, allowing the force the atoms have experienced to be measured very precisely. Atom interferometry measurements looking for chameleons have reached a sensitivity of $10^{-6} g$ (where $g$ is the acceleration due to free fall at the surface of the earth) and are forecast to reach $10^{-9} g$ \cite{Hamilton:2015zga,Elder:2016yxm}.

For chameleon screening two properties make atom interferometry particularly powerful.  Firstly, because the atoms are so small they are unscreened and so the chameleon force is less suppressed than it would be in a comparable macroscopic fifth force experiment.  Secondly the walls of a vacuum chamber are sufficiently thick that they screen the interior of the vacuum chamber from any chameleon gradients or fluctuations in the exterior.  This simplifies the computation of the chameleon forces in the experiment, but does have the consequence that the source mass must be placed inside the vacuum chamber, unlike many other tests of gravity performed with atom interferometry. 

Do these same advantages also apply to symmetron screening?  Assuming the walls of the vacuum chamber have a density of $\rho_{\rm wall}\sim \mbox{g/cm}^3$ then the field can reach $\phi=0$ and restore the symmetry inside the walls if the thickness of the walls is greater than $\sim 1/m_{\rm wall}$.  We will see shortly that atom interferometry experiments constrain a fairly narrow region in $\mu$ around $\mu \sim 10^{-4}\mbox{ eV}$ and coupling strengths in the range $10^{-4} \mbox{ GeV}< M < 10^{4} \mbox{ GeV}$. 
In this region of the parameter space the Compton wavelength of the field in the symmetry restored vacuum in the wall has a maximum value of $1/m_{\rm wall} \sim 1\mbox{ mm}$.  Therefore we should expect the symmetry to be restored in the wall in the region of parameter space we consider, and so the interior is effectively decoupled from the behaviour of the symmetron in the exterior.  

For a compact object to be screened from the symmetron force we need both $\rho_{\rm in}/M^2 >\mu^2$ and $m_{\rm in}R\gg1$. The first condition is actually {\it easier } to satisfy for an atomic nucleus than for a macroscopic test mass, because the nuclear density is much higher than the density of, for example, silicon. The second condition is harder to satisfy for atoms than macroscopic masses because of the small size of the atomic nuclei.  It is therefore not always the case that atoms make better probes of the symmetron field than macroscopic objects do, however they will be sensitive in some region of the parameter space which we will now determine. 

\begin{figure}
\centering
\includegraphics[scale=0.8]{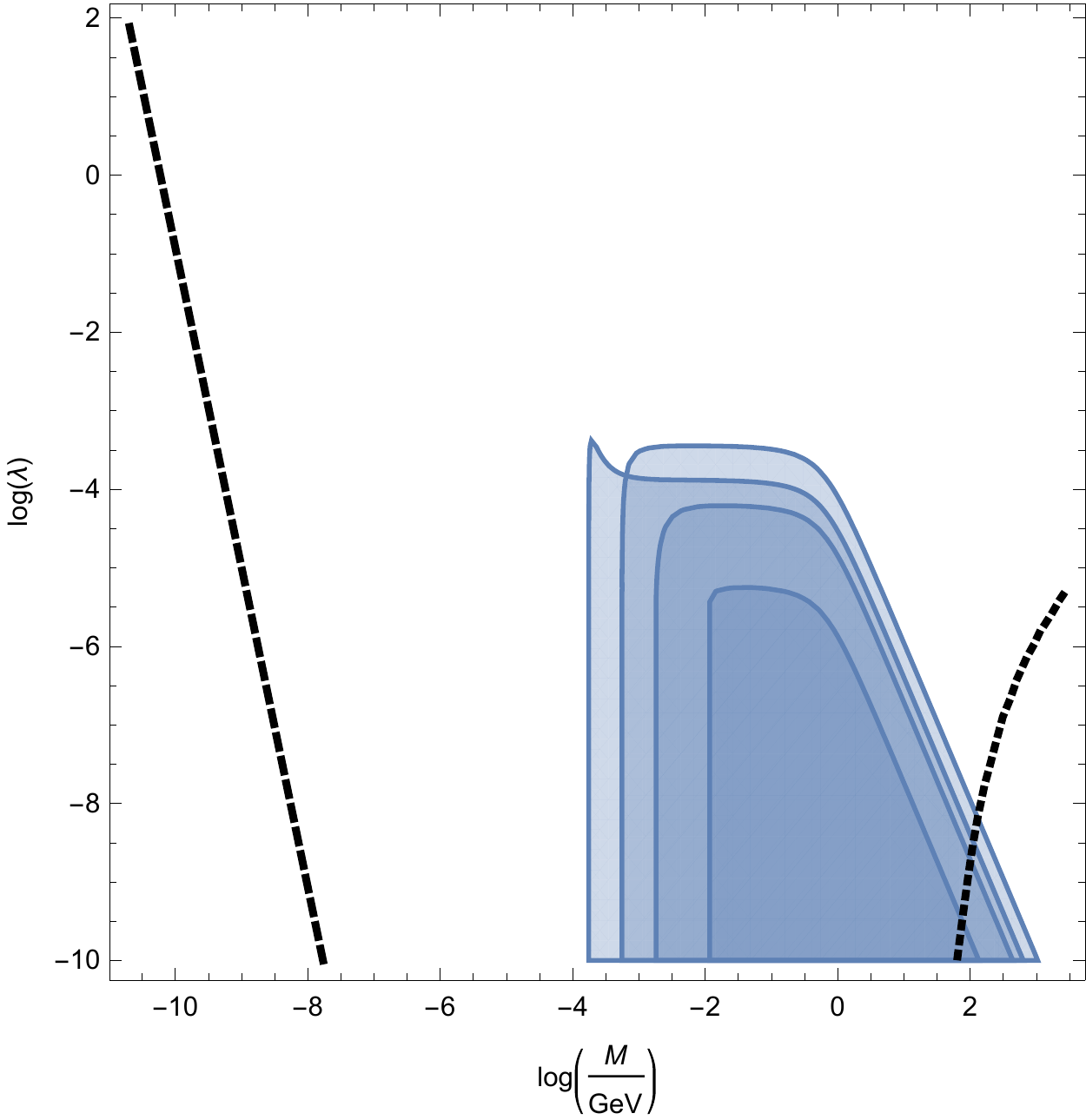}
\caption{\label{fig:constraints} Constraints on the symmetron parameters from the atom interferometry experiment of \cite{Hamilton:2015zga}. The excluded regions are shaded blue.  The different regions are for different values of $\mu$; from left to right $\mu= 10^{-4}\mbox{ eV}$, $\mu= 10^{-4.5}\mbox{ eV}$, $\mu= 10^{-5}\mbox{ eV}$, $\mu= 10^{-5.5}\mbox{ eV}$.  The black dashed line on the left   shows constraints from observations of exo-planets (points to the left of the line are excluded), with $\mu \rightarrow 0$ the most constraining choice \cite{Santos:2016rdg}.   The black dotted line on the right shows the constraints from torsion pendulum experiments (points to the right of the line are excluded) with $\mu=10^{-4} \mbox{ eV}$ chosen for reference  \cite{Upadhye:2012rc}.}
\end{figure}

We apply the results of reference \cite{Hamilton:2015zga}, which measures the acceleration of cold caesium 133 atoms.  The atoms were held $8.8\mbox{ mm}$ away from an aluminium sphere of radius $9.5\mbox{ mm}$. The experiment was performed in a vacuum chamber of radius $5 \mbox{ cm}$ and pressure $6 \times 10^{-10}\mbox{ Torr}$. No anomalous acceleration of the atoms is measured, restricting the acceleration due to the symmetron field to satisfy  $a< 6.8 \times 10^{-6}\mbox{ m/s}^2$. The constraints that this places on the symmetron parameter space can be seen in Figure \ref{fig:constraints}.

Constraints are restricted to a narrow range of the mass parameter $\mu$, around $\sim 10^{-4}, 10^{-5}\mbox{ eV}$.  For smaller $\mu$ the Compton wavelength of the symmetron in the vacuum is larger than the size of the vacuum chamber and so the field cannot vary its value over the scale of the experiment.  For larger $\mu$ the Compton wavelength of the symmetron in vacuum is so small that the Yukawa term exponentially suppresses the force. The peak in the $\mu= 10^{-4}\mbox{ eV}$ plot occurs because there is a value of $M$ for which the Compton wavelength of the field in vacuum exactly matches the distance between the aluminium sphere and the atoms.

We see that, whilst the range of $\mu$ values that are accessible is relatively constrained, where atom interferometry experiments do give constraints they explore a region of parameter space that is inaccessible to other experiments and observations.

\subsection{Other Experiments with Unscreened Test Particles}
Atoms are not the only objects that can be unscreened in a laboratory vacuum.  Experiments that measure forces on neutrons \cite{Brax:2011hb,Ivanov:2012cb,Brax:2013cfa,Jenke:2014yel,Lemmel:2015kwa,Li:2016tux} and on silicon microspheres \cite{Rider:2016xaq} have also been shown to be sensitive to chameleon forces, precisely because the test particles are sufficiently small that they are not screened.  However, these have not yet reached the sensitivity of the atom interferometry experiments and so do not provide better constraints on symmetron models than those presented in Figure \ref{fig:constraints}.  We note, however, that silicone microspheres have a lower average density than neutrons and atomic nuclei, and so, if the sensitivity can be improved, they may provide the best prospect for searching for symmetron forces. 

\section{Domain Walls}
\label{sec:domain}
Symmetron fields open another possibility for laboratory searches that is not present for the chameleon model. Since the symmetron effective potential has two minima, as gas is pumped from the vacuum chamber the field could settle in either minimum with equal probability, and there is no reason for different regions of the chamber to all settle into the same one; if the Compton wavelength is comparable to or smaller than the size of the vacuum chamber then it is possible for a domain wall or a network of domain walls to form.

  If we approximate the wall as being straight and static, then its field profile is
\begin{equation}
\phi(z)=\phi_0\tanh\left(\frac{\tilde{\mu}z}{\sqrt{2}}\right)\;,
\label{eq:soliton}
\end{equation}
where 
\begin{equation} \label{eq_muTilde}
\tilde{\mu}^2:=\mu^2-\frac{\rho}{M^2}\;,
\end{equation}
and $\phi_0 = \frac{\tilde{\mu}}{\sqrt{\lambda}}$, meaning that the thickness of the wall is $\sim 1/\tilde{\mu}$, and its tension is $= 4\tilde{mu}^3/(3\lambda)$ \cite{Vilenkin:2000jqa}.  

Taking into account that the atom may be screened, the acceleration experienced by  atom moving in the neighbourhood of a domain wall  is 
\begin{equation}
\vec{a} = 4 \pi \lambda_{\rm atom} (1+m_{\rm out}R_{\rm atom}) \nabla \phi\;,
\end{equation}
where we should remember that $\lambda_{\rm atom}$ depends on the background field value $\phi_{\rm out}$ which in this case should be replaced with the domain wall field profile $\phi$ evaluated at the position of the atom.

The maximum acceleration that an atom may experience in such a situation is roughly proportional to $|\phi\nabla\phi|$. We can find an approximation for this by assuming the domain walls are small compared to the radius of the chamber  such that we can use the planar solution for a domain wall given above in equation (\ref{eq:soliton}). We find that this maximum acceleration $a_\phi\approx\frac{\phi\nabla\phi}{M^2}\approx 0.27\frac{\tilde{\mu}^3}{\lambda M^2}$, which   is always much less than $10^{-10}\;g$  within the parameter space we have examined. This means that any domain walls that form will have a negligible effect on searches for symmetron fifth forces between atoms and source masses in the vacuum chamber, and that sensitivity must be improved if we are to detect the forces due to the domain wall directly.

\begin{figure}
\centering
\includegraphics[scale=0.65]{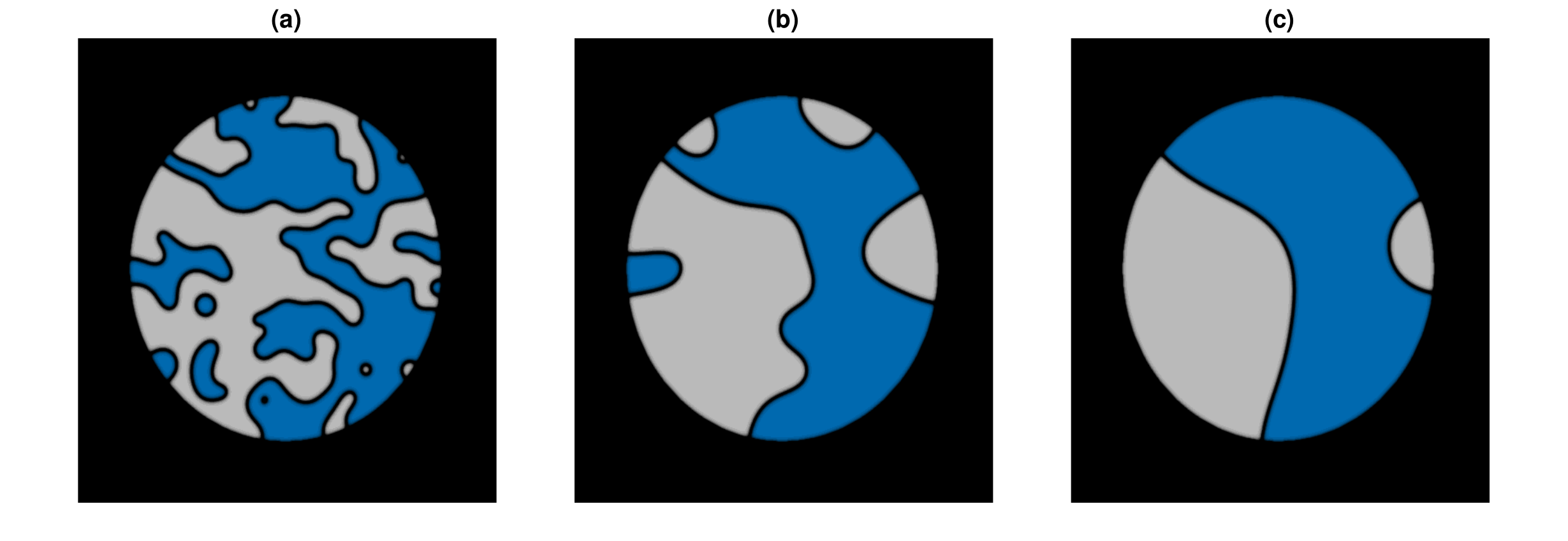}
\caption{\label{fig:network} A two dimensional domain wall network inside a circular cavity. The colours indicate the value of the scalar field.  Blue and grey regions represent the positive and negative symmetry broken vacua respectively, and in the black regions the symmetry is restored at $\phi=0$. The system evolves in time from left to right.  This simulation was performed with unphysical values for the symmetron parameters due to numerical limitations.  }
\end{figure}

Of course, depending on the correlation length more than one domain wall can form creating a network. The symmetry is restored in the walls of the vacuum chamber, and in the core of the domain wall, so from the point of view of the field the walls can be viewed as a fixed sphere of $\phi=0$ surrounding the domain wall network.  We know from cosmological studies \cite{Vilenkin:2000jqa} that networks of  domain walls want to evolve towards the configuration with the minimum wall length.  Therefore we can assume that the network inside the vacuum chamber is not stable, and the domain walls will straighten, and merge with one another and with the walls of the vacuum chamber.  It is therefore reasonable to expect that the end point of this evolution will be the vacuum chamber entirely filled with one domain and no domain walls are present.
 An example of such an evolution is shown in Figure \ref{fig:network}.  This figure was constructed using two-dimensional numerical simulation with unphysical values for the symmetron parameters, and so should be considered only as an example of what kind of evolution is possible.  We leave a full numerical simulation for future work. 

 It remains to be determined how long such an evolution takes. We can use some results from cosmological studies of domain walls as a guideline of what to expect.  In a true vacuum it is expected that the domain walls move with relativistic velocities.  If this were the case in our vacuum chamber the walls would exist for an undetectably short period of time.  However, this motion can be slowed down by friction if the walls interact with a surrounding particle bath.  The force per unit area on the wall can be approximated by \cite{Kibble:1976sj}
\begin{equation}
F_{\rm friction} \sim N n T v\;,
\end{equation}
where $N$ is the number of light  particles interacting with the domain wall, $n$ is the number density of these particles, $T$ is their temperature and $v$ the velocity of the wall compared to the background. Taking the number density corresponding to the hydrogen gas pressure in the atom interferometry experiment described above, which is performed at room temperature, we find
\begin{equation}
F_{\rm friction} \sim v \times 1.3 \times 10^{-45} \mbox{ GeV}^4\;.
\end{equation}
This frictional force is comparable to the domain wall tension if $F_{\rm friction} \sim 4 \mu^3/(3 \lambda R)$, where $R$ is the mean curvature radius. From this we can deduce that the strings will move non-relativistically if
\begin{equation}
0.02 \left(\frac{\mbox{cm}}{R}\right) \ll \lambda\;.
\end{equation}
This suggests that at least for some values of $\lambda$, the domain walls could be long lived inside the vacuum chamber.  As mentioned above, a full numerical study of the evolution of the domain walls in a vacuum chamber remains a topic for future work.

Searches for the forces due to domain walls are not the most sensitive way to search for symmetron fields, although they do have the technical advantage that they do not rely on the presence of a source mass that can be moved inside the vacuum chamber.  However, they only occur in theories of screening, such as the symmetron, which undergo a symmetry breaking transition.  If a fifth force with screening is detected in an upcoming experiment, the presence or absence of a network of domain walls in the experiment could be used to discriminate between models.

\section{Conclusions}\label{sec:conclusions}

We have shown that symmetron fifth forces, inspired by theories of dark energy, can be constrained by terrestrial experiments using cold atoms.  The constraints we have found in Figure \ref{fig:constraints}, are particularly interesting as they fill a previously empty region of parameter space between the constraints coming from E\"{o}t-Wash experiments, and those coming from observations of exo-planets. 

We have also discussed the possibility that symmetron domain walls may form in the vacuum chamber, leading to the atoms experiencing a fifth force without the need to place a source mass inside the vacuum chamber. Whilst we find that the accelerations experienced by the atoms are smaller than the sensitivity of current experiments, they are not so small that it would be impossible to detect them in the future.  Additionally, as the domain walls only form for symmetron models, if a screened fifth force is ever detected in a terrestrial experiment the presence or absence of these domain walls would provide a way to discriminate between different models of screening. 

\subsection*{}
In the final stages of writing this article it has come to our attention that Brax and Davis have derived the same constraints on the symmetron model using the tomographic model of screening \cite{A&P}.

\section*{Acknowledgements}
We would like to thank Ed Copeland for useful discussions during the completion of this work. CB is supported by a Royal Society University Research Fellowship. JS is supported by the Royal Society

%
%
\bibliography{draft}
\bibliographystyle{unsrt}

\end{document}